\documentstyle[graphicx]{article}

\begin{document}
\title{Frequency analysis of tick quotes on the foreign exchange market\\
and agent-based modeling: A spectral distance approach}
\author{Aki-Hiro Sato \\ Department of Applied Mathematics and Physics, \\ 
Graduate School of Informatics, Kyoto University, \\
Kyoto 606-8501, Japan.} 
\maketitle
\begin{abstract}
High-frequency financial data of the foreign exchange market (EUR/CHF,
 EUR/GBP, EUR/JPY, EUR/NOK, EUR/SEK, EUR/USD, NZD/USD, USD/CAD, USD/CHF,
 USD/JPY, USD/NOK, and USD/SEK) are analyzed by utilizing the
 Kullback-Leibler divergence between two normalized spectrograms of 
 the tick frequency and the generalized Jensen-Shannon divergence among
 them. The temporal structure variations of the
 similarity between currency pairs is detected and characterized. A
 simple agent-based model in which $N$ market participants exchange $M$
 currency pairs is proposed. The equation for the tick frequency is
 approximately derived theoretically. Based on the analysis of this
 model, the spectral distance of the tick frequency is associated with
 the similarity of the behavior (perception and decision) of the market
 participants in exchanging these currency pairs.
\end{abstract}
\section{Introduction}
\label{sec:introduction}
The recent accumulation of high-frequency financial data due to
the development and spread of information and communications technology has 
sparked interest in financial markets~\cite{Mantegna:00,Dacorogna:00,Strozzi:02,Mizuno:03,Ohnishi:04,Petroni:04,Aiba:04,Suzuki:04,Wang:06,Kiyono:06}.
Many researchers expect new findings and insights into
the worlds of both finance and physics. Since the financial markets are
complex systems that consist of several agents that interact with one
another, an enormous amount of data must be treated in order to describe
and understand them at the microscopic level. Therefore, it is important
to find adequate variables or relevant quantities to describe their
properties~\cite{Haken:88}. Since a macroscopic description allows
information with global properties to be compressed, if the adequate
macroscopic quantities can be determined, then relationships can be
established among various macroscopic quantities and a deeper
understanding of the system can be obtained. 

On the other hand, agent-based models as complex systems are attracting 
significant interest across a broad range of disciplines. Several
agent-based models have been proposed to explain the behavior of
financial markets during the last
decade~\cite{Aoki:96,Lux:99,Challet:00,Kaizoji:00,Krawiecki:03,Tanaka:03,Sato:04}. Agent-based models are expected to provide an 
alternative to phenomenological models that mimic market price
fluctuations. Specifically, it seems to be worth considering the
explanation capability of the agent-based models for causality from a 
microscopic point of view.

In a previous study, the tick frequency, which is defined as the number
of tick quotations per unit time, was reported to exhibit periodic
motions due to the collective behavior of the market
participants~\cite{Sato:06-a,Sato:06-b}. As a result, the tick frequency
appears to be an important representative quantity in the financial
market. Moreover, it has been reported that it is possible to detect the
dynamic structure of the foreign exchange market by using the spectral
distance defined by the Kullback-Leibler
divergence~\cite{Sato:06-c}. The spectral distance of the tick frequency
is one possibility for macroscopically describing the relationship among
market participants in the financial market.

In the present study, the meaning of the spectral distance of the tick
frequency is discussed, starting from the microscopic description with
the agent-based model of a financial market. First, definitions and the
results of the spectral distance of the tick frequency are presented.
Next, a model that consists of $N$ market participants who choose
their action among three kinds of investment attitudes in order to exchange $M$
currency pairs is considered. In this model, the heterogeneous agents
perceive information in the environment, which is separated into exogenous
factors (news about events) and endogenous factors (news about market
fluctuations), and decide their actions based on these factors. There are two
thresholds by which to select their actions among three kinds of investment
attitudes (buying, selling, and waiting). Analysis of this model
indicates that the spectral distance of the tick frequency is equivalent
to the difference among behavioral parameters of market participants who
exchange these currency pairs.

This remainder of this paper is organized as follows. In Section \ref{sec:spectral-analysis}, the tick frequencies of 12 currency pairs in the foreign 
exchange market are analyzed with the spectral distance measured by 
the Kullback-Leibler divergence and the Jensen-Shannon divergence of 
the normalized power spectra. In Section \ref{sec:agent-based-model}, an
agent-based model in which $N$ market participants deal with $M$ currency
pairs is proposed. In Section \ref{sec:discussion}, based on the agent-based
model, the equation for the tick frequency is approximately derived, and
the relationship between the spectral distance of the tick frequency and 
the behavioral parameters of market participants is
discussed. Section \ref{sec:conclusions} is devoted to concluding
remarks.

\section{Frequency analysis}
\label{sec:spectral-analysis}
\subsection{Data}
The foreign currency market data of United States Dollar
(USD), Euro (EUR), Switzerland Francs (CHF), Great Britain Pounds
(GBP), Norwegian Krone (NOK), Swedish Krona (SEK), Canadian Dollars (CAD), New
Zealand Dollars (NZD), and Japanese Yen (JPY), as provided by CQG
Inc., were investigated~\cite{CQG}. The data include two quote rates,
namely, the bid rate and the ask rate, with a resolution of one
minute. The bid and ask rates are the prices at which bank traders are
willing to buy and sell a unit of currency. All traders in the foreign
exchange market have a rule to quotes both rates at the same time
(two-way quotation). Generally, the ask rate is higher than the bid
rate, and the difference between the bid rate and the ask rate is called
the bid-ask spread.  

The data investigated in this article are from two databases. The first
includes 12 currency pairs, EUR/CHF, EUR/GBP, EUR/JPY, EUR/NOK, EUR/SEK, 
EUR/USD, NZD/USD, USD/CAD, USD/CHF, USD/JPY, USD/NOK, and USD/SEK, during
the period from the 1st to the 29th of September 2000. The other
includes three currency pairs, EUR/USD, EUR/JPY, and USD/JPY, during the
period from the 4th of January 1999 to the 31st of December 2004. The
data start at 17:00 (CST) on Sunday, and finish at 16:59 (CST) on
Friday. The foreign exchange market is open for 24 hours on weekdays. On
Saturdays and Sundays, there are no ticks on the data set because most
banks are closed.

\subsection{Methods}
The tick frequency is defined by counting the number of times that bank traders
quote the bid and ask rates per unit time. According to this definition 
a currency pair having a high (low) quote frequency indicates activity
(inactivity). Since bank traders usually quote both bid and ask rates at
the same time, it is sufficient to count only the bid or the ask quotation. 
Here, the tick frequency is defined as the number of ask quotes per unit time,
\begin{equation}
A(k) = \frac{1}{\Delta t}C(k\Delta t;(k+1)\Delta t), \quad (k=0,1,\ldots)
\end{equation}
where $C(t_1;t_2)$ is the number of the ask quotes from $t_1$ to $t_2$,
and $\Delta t$ denotes the sampling period, and $\Delta t=1$ minutes
throughout this analysis.

The spectrogram is estimated with a discrete Fourier transform for a time
series multiplied by the Hanning window. Let $A(k) \quad
(k=0,1,2,\ldots)$ be the time series of tick frequencies. The spectrogram with 
the Hanning window represented as
\begin{equation}
w(k) = \frac{1}{2}\Bigl(1-\cos(\frac{2\pi k}{L-1})\Bigr),
\end{equation}
is defined as 
\begin{eqnarray}
P(n,T) &=& \Bigl|\sum_{k=0}^{L-1}w(k)A(k+T)e^{-2\pi \mbox{i}
 k \frac{n}{L}}\Bigr|^2, 
\end{eqnarray}
where $T$ denotes the representative time to localize the data by the window
function having the range of $L\Delta t$. Since the Nyquist critical
frequency is given by $f_c = 1/(2\Delta t) = 0.5$ [1/min], the behavior of
the market participants can be detected with a resolution of 2 [min]. 

For the purpose of quantifying the similarity between the tick
frequencies, the Kullback-Leibler (KL) divergence method is applied
between normalized spectrograms of the tick frequency~\cite{Sato:06-a}. 
The KL is defined as a functional of two normalized positive
functions~\cite{Amari:00}. In order to apply the KL method to the
spectrogram, the normalized spectrogram is introduce as follows:
\begin{equation}
p(n,T) = \frac{P(n,T)}{\sum_{n=1}^{L-1}P(n,T)}.
\end{equation}
The KL between the spectrograms without the direct current component 
for the $i$-th currency pair, $p_i(n,T)$, and that for the $j$-th
currency pair, $p_j(n,T)$, is defined as 
\begin{equation}
K_{ij}(T) = \sum_{n=1}^{L-1}p_i(n,T)\log\frac{p_i(n,T)}{p_j(n,T)}.
\end{equation}
Based on its definition, the KL is always non-negative,
\begin{equation}
K_{ij}(T) \geq 0,
\end{equation}
with $K_{ij}(T) = 0$ if and only if $p_i(n,T) = p_j(n,T)$. Note that
$K_{ij}(T)$ is an asymmetric dissimilarity matrix, which is satisfied by 
\begin{equation}
K_{ij}(T) \neq K_{ji}(T), K_{ii}(T) = 0.
\end{equation}
Generally, the asymmetric matrix is separated into a symmetric matrix and
an asymmetric matrix,
\begin{equation}
K_{ij}(T) = J_{ij}(T) + I_{ij}(T),
\end{equation}
where $J_{ij}(T) = (K_{ij}(T)+K_{ji}(T))/2$, $I_{ij}(T) =
(K_{ij}(T)-K_{ji}(T))/2$. Specifically, $J_{ij}(T)$ is called the  
symmetrical Kullback-Leibler distance (SKL) and is defined as follows:
\begin{equation}
J_{ij}(T) =
 \frac{1}{2}\sum_{n=1}^{L-1}(p_i(n,T)-p_j(n,T))
\log\frac{p_i(n,T)}{p_j(n,T)},
\end{equation}
where $J_{ij}(T)=J_{ji}(T)$, $J_{ii}(T)=0$, and $J_{ij}(T)=0$ if and only if
$p_i(n,T)=p_j(n,T)$.

As an alternative symmetric divergence, the Jensen-Shannon
divergence (JS)~\cite{Lin:91} is defined as follows:
\begin{equation}
JS_{ij}^\pi(T) = 
  H(\pi_ip_i+\pi_jp_j,T)-\pi_iH(p_i,T)-\pi_jH(p_j,T),
\end{equation}
where $\pi_i,\pi_j>0$, $\pi_i+\pi_j=1$ are the {\it a priori} probability
for $p_i$ and $p_j$, and $H(p,T)$ is the Shannon entropy, which is
defined as $H(p,T) = -\sum_{n=1}^{L-1}p(n,T)\log p(n,T)$. This
divergence possesses symmetric and non-negative features:
$JS_{ij}^\pi(T) = JS_{ji}^\pi(T)$, $JS_{ij}^\pi(T)\geq 0$, and
$JS_{ij}^{\pi}(T)=0$ if and only if $p_i(n,T)=p_j(n,T)$. Moreover, it is
possible to calculate the total similarity of the market using the
generalized Jensen-Shannon divergence (GJS),
\begin{equation}
JS^\pi(p_1,p_2,\ldots,p_M) = H\Bigl(\sum_{i=1}^M\pi_ip_i\Bigr) - 
\sum_{i=1}^M\pi_iH(p_i).
\end{equation}
The GJS is non-negative, i.e., $JS^\pi(p_1,p_2,\ldots,p_M)\geq 0$ and
$JS^\pi(p_1,p_2,\ldots,p_M)=0$, if and only if
$p_1(n,T)=p_2(n,T)=\cdots=p_M(n,T)$.

\subsection{Results}
\begin{figure}[hbt]
\centering
\includegraphics[scale=0.38]{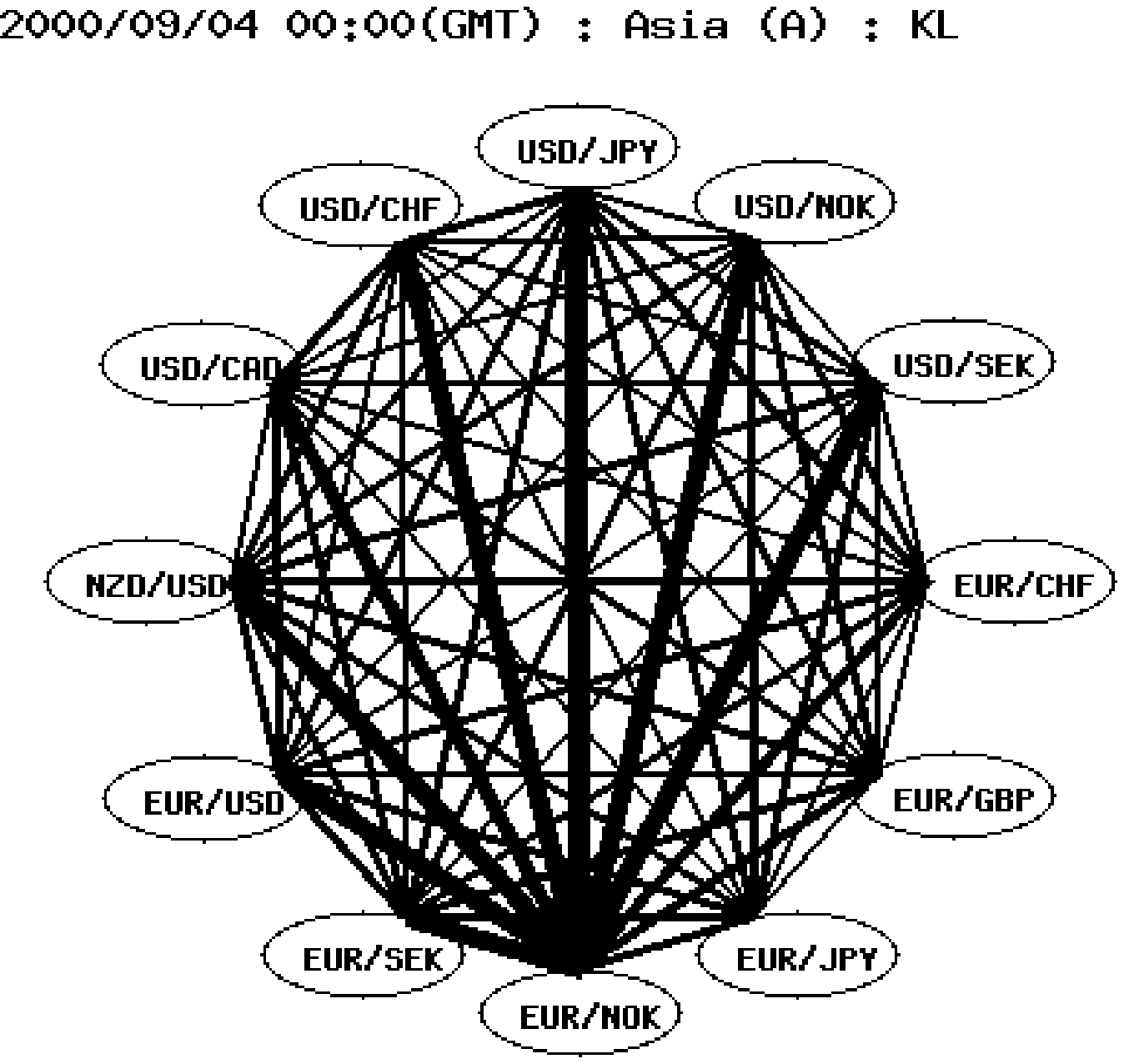}
\includegraphics[scale=0.38]{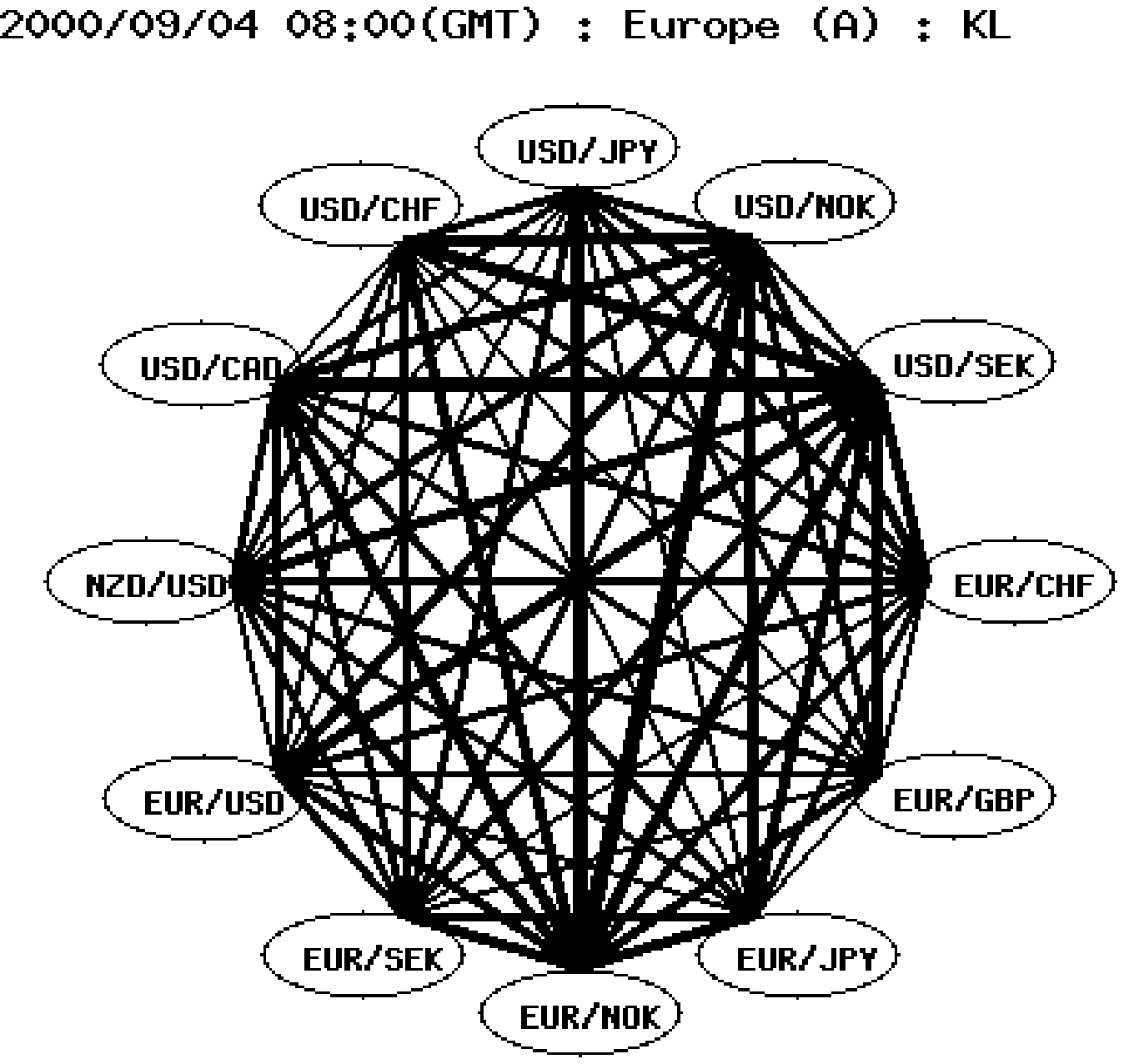}
\includegraphics[scale=0.38]{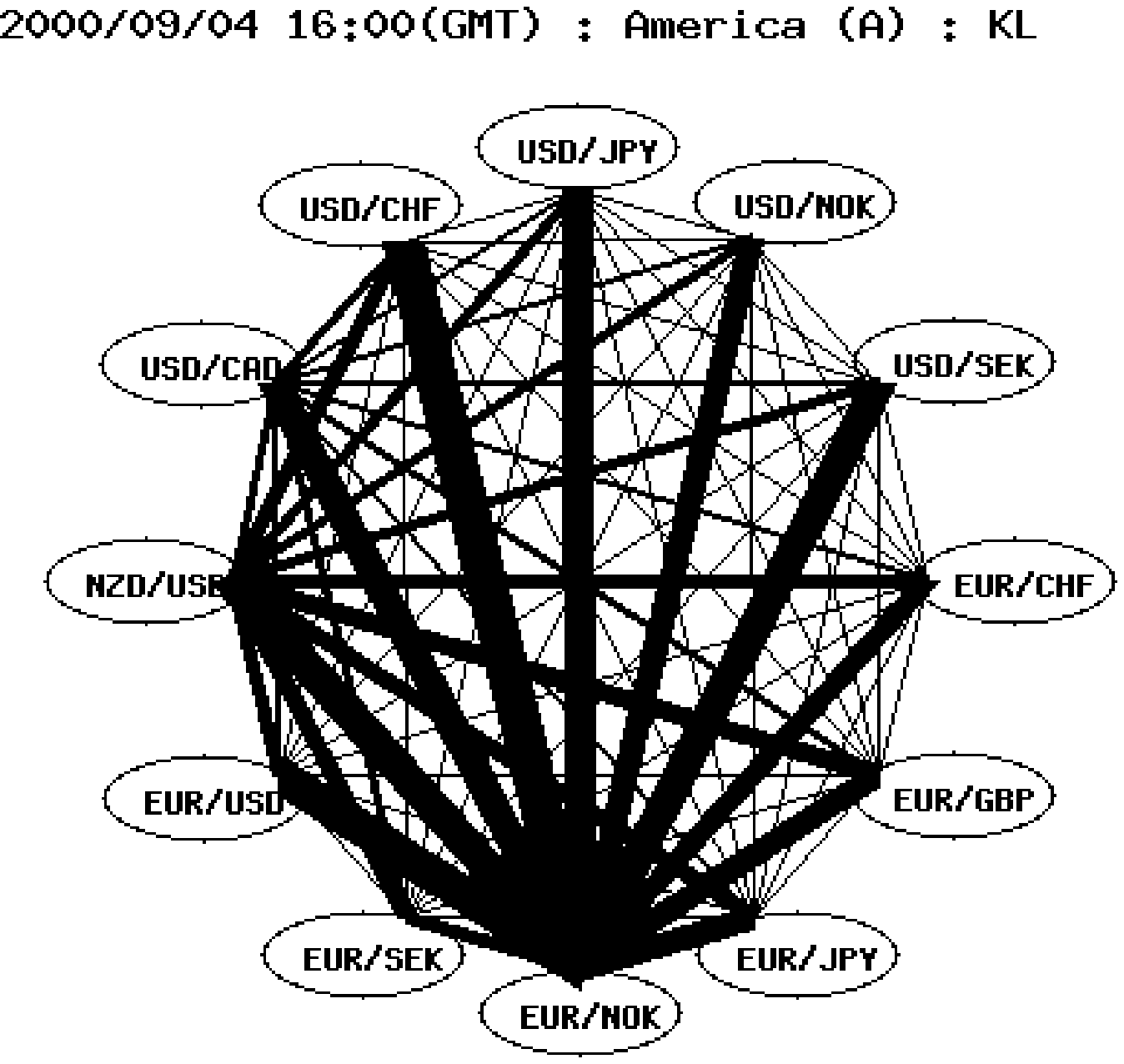}
\caption{The symmetrical Kullback-Leibler distance among 12 currency pairs 
for the Asian time zone ($T=$0:00 (UTC+1)) (left), the European time zone
 ($T=$8:00 (UTC+1)) (center), and the American time zone ($T=$16:00
 (GMT)) (right) on the 4th of September 2000 for $N=480$ (8 hours). The
 thin/thick lines between the $i$-th currency pair and the $j$-th
 currency pair represent the similarity/dissimilarity between them,
 $J_{ij}(T)$.}
\label{fig:network}
\end{figure}

Figure \ref{fig:network} shows the SKL among 12 currency pairs for the
Asian time zone ($T=$0:00 (UTC+1)), the European time zone ($T=$8:00
(UTC+1)), and the American time zone ($T=$16:00 (UTC+1)) as fully
connected networks.

The patterns in Fig. \ref{fig:network} indicate the existence of similar
currency pairs and dissimilar currency pairs at each time zone. The
similarity and dissimilarity between currency pairs varies
temporally. Furthermore, the variations of the network structure are
slow. The results obtained using the JS provide are similar to those
obtained using the SKL. 

\begin{figure}[hbt]
\centering
\includegraphics[scale=0.35]{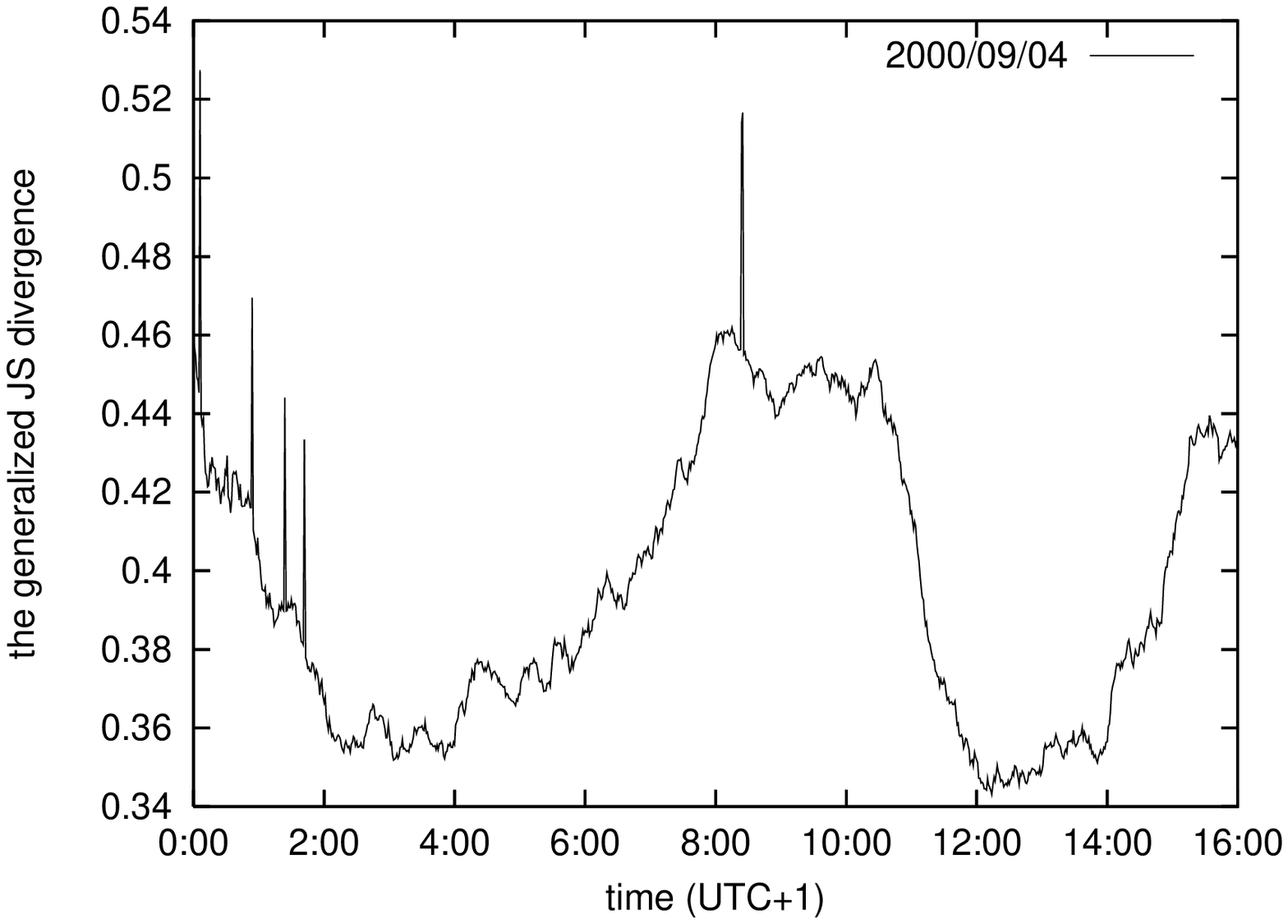}
\caption{Time series of the generalized Jensen-Shannon divergence among 
12 currency pairs with $L=480$ [min] on the 4th of September 2000.}
\label{fig:JS12}
\end{figure}

As shown in Fig. \ref{fig:JS12} the GJS among 12 currency pairs under
the uniform condition $\pi_i = 1/12$ varies periodically. This periodic
variation is thought to be due to the life cycle of the market
participants. Specifically, the GJS becomes low at the time between time
zones. Outliers, which look like pulses, occasionally occur, indicating
drastic changes in the similarity structure at a specific time. 

\begin{figure}[hbt]
\begin{center}
\includegraphics[scale=0.3]{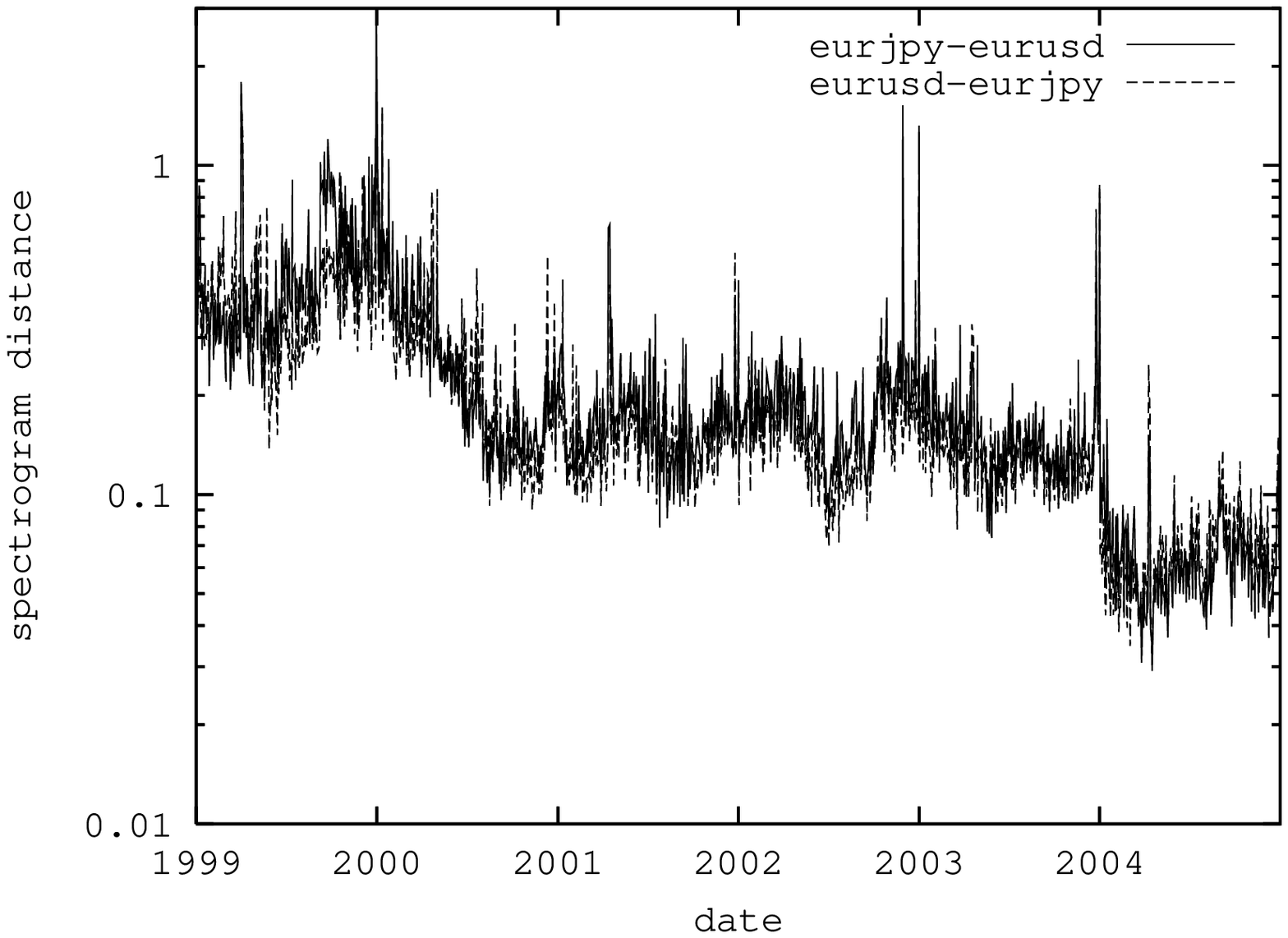}(a)
\includegraphics[scale=0.3]{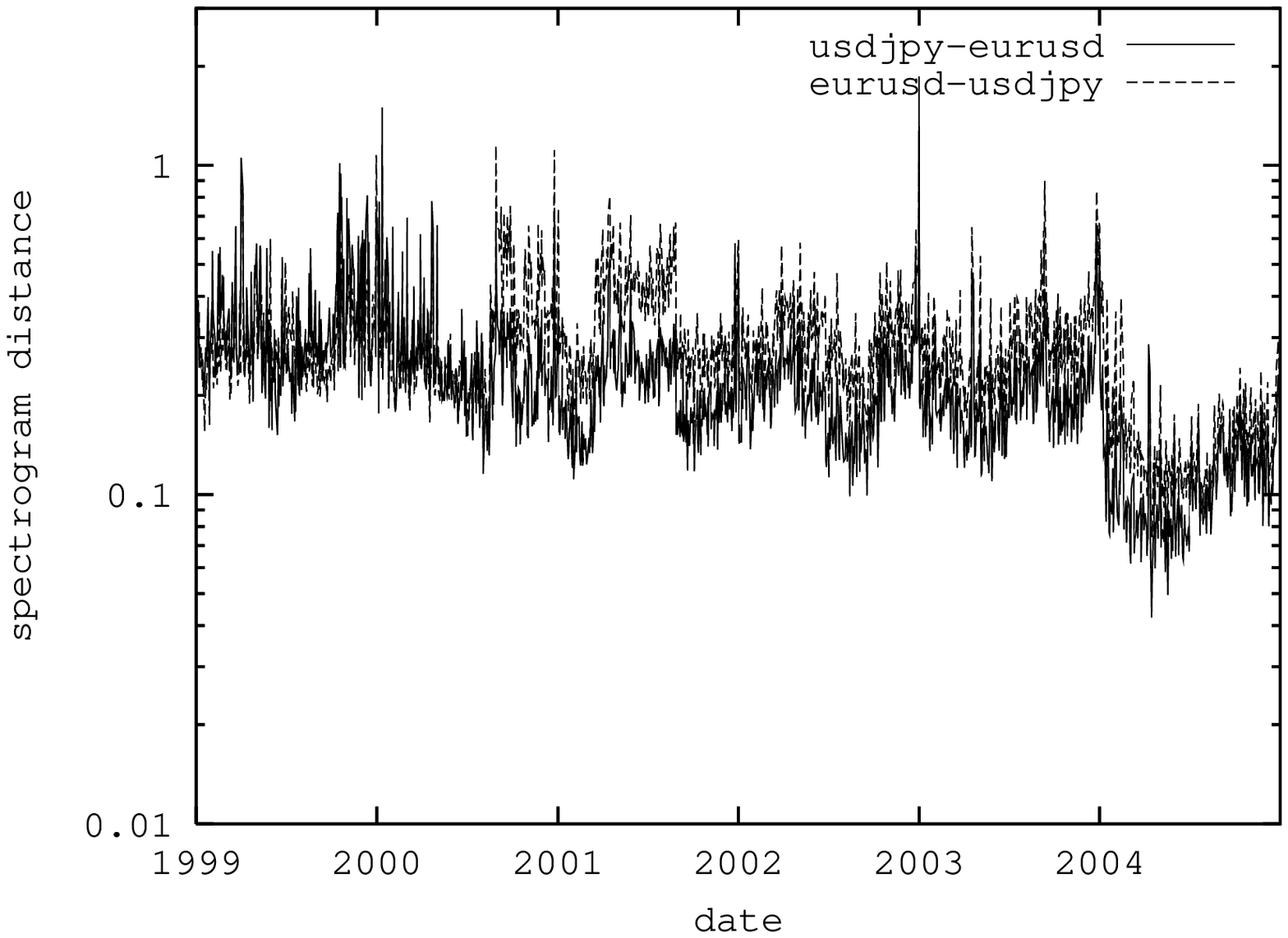}(b)
\includegraphics[scale=0.3]{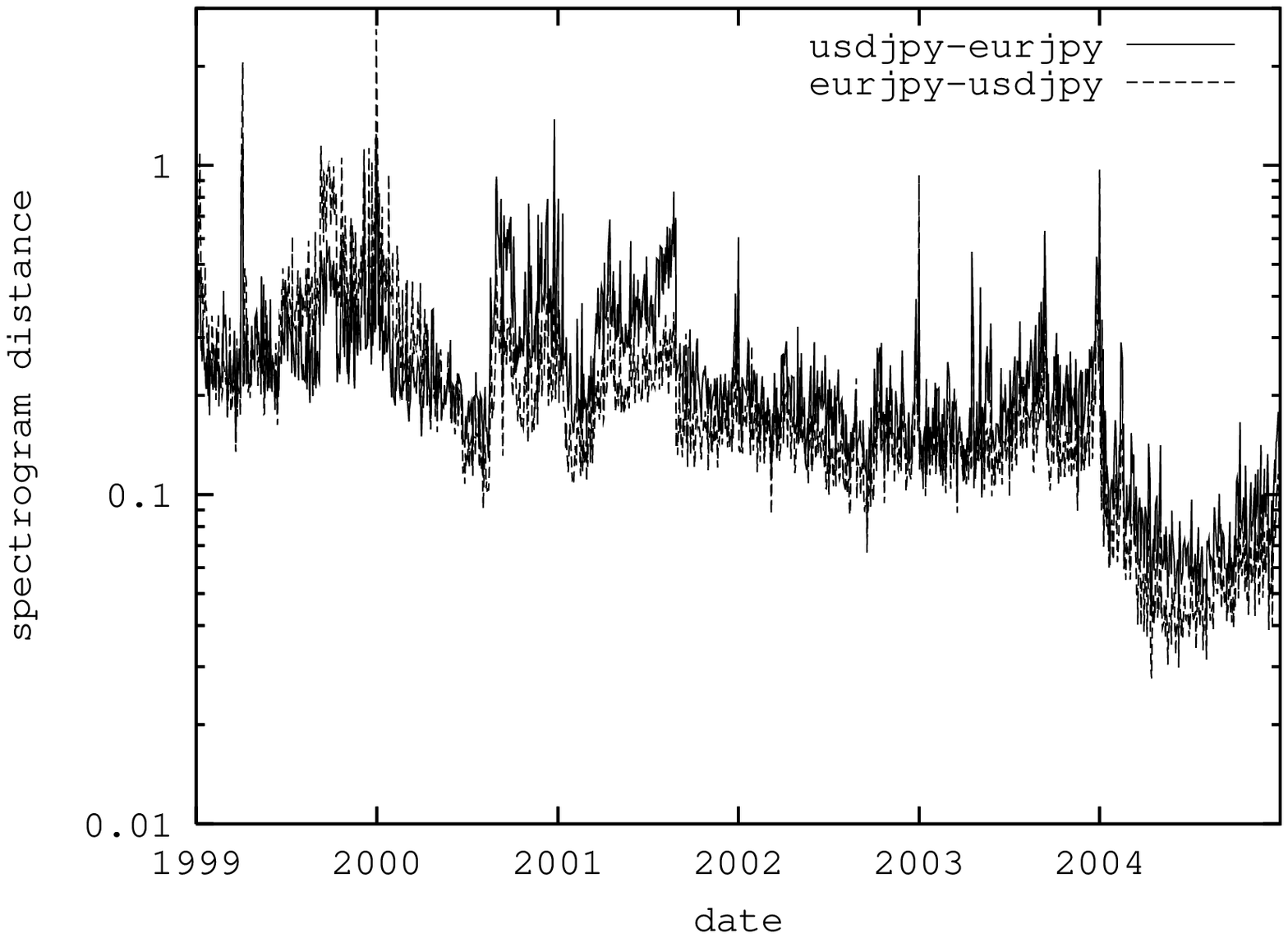}(c)
\end{center}
\caption{The Kullback-Leibler divergence of spectrograms between the
 EUR/USD and the EUR/JPY (a), between the EUR/USD and the USD/JPY (b),
 and between the EUR/USD and the EUR/JPY (c) for the period between the 4th
of January 1999 and the 31st of December 2004.}
\label{fig:KL}
\end{figure}

The results of the SKL divergences among the EUR/USD, the USD/JPY, and
the EUR/JPY during the period from the 4th of January 1999 to the 31st
of December 2004 are shown in Fig. \ref{fig:KL}. The SKL divergences
decrease yearly and have high values around the first day of each year.

\begin{figure}[hbt]
\begin{center}
\includegraphics[scale=0.35]{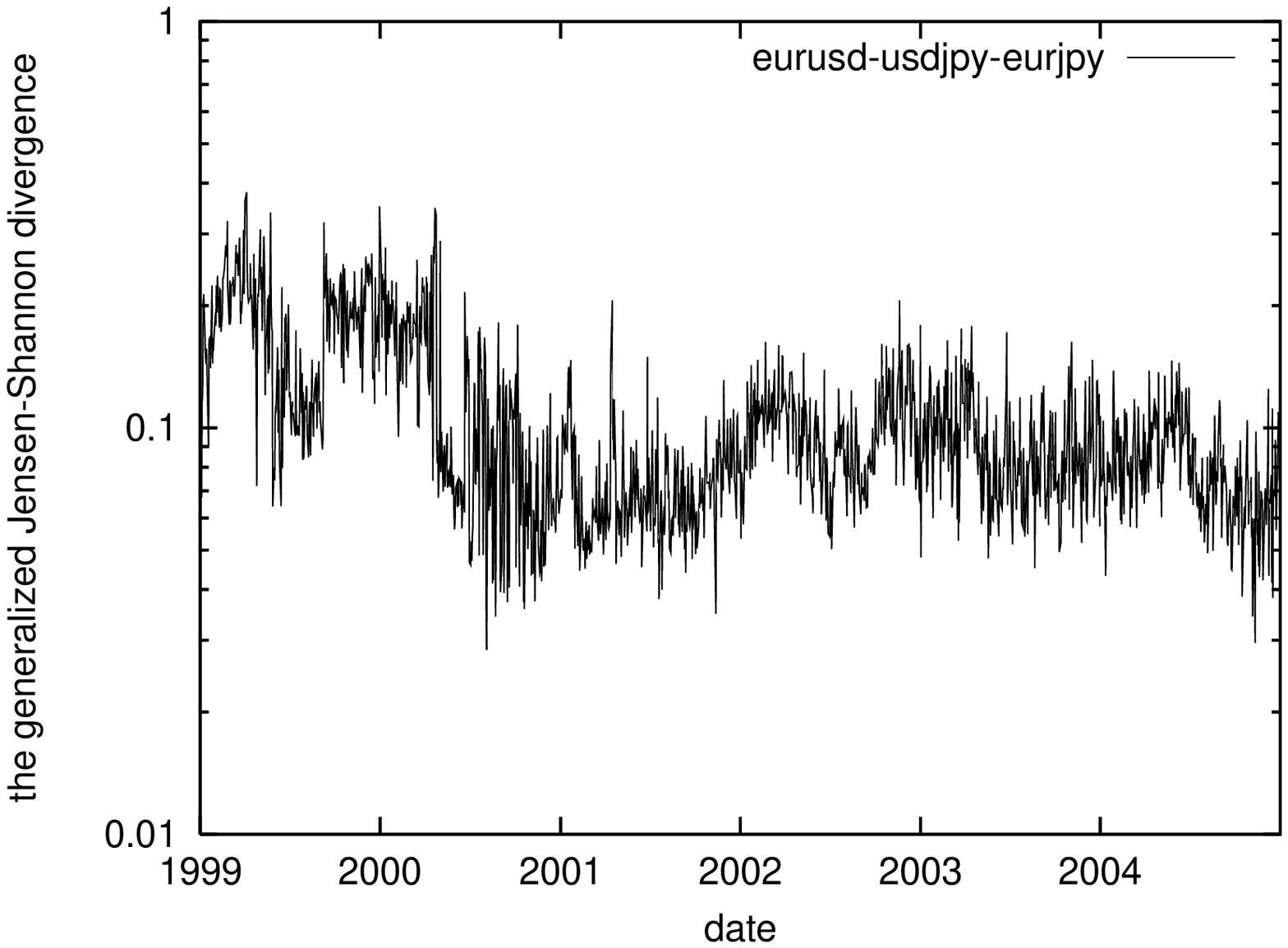}
\end{center}
\caption{The generalized Jensen-Shannon divergence of spectrograms
among the EUR/USD, USD/JPY, and EUR/JPY for the period between the 4th of January
1999 and the 31st of December 2004.}
\label{fig:JS3}
\end{figure}

The results of the GJS among the EUR/USD, the USD/JPY, and the EUR/JPY 
during the period from the 4th of January 1999 to the 31st of December
2004 are shown in Fig. \ref{fig:JS3}. The GJS values in 1999 are
relatively higher than those after 2001. For the period from the middle
of 1999 to the middle of 2000, the GJS values remain high. The GJS
values decrease from the middle of 2000 and tend to be less than
0.1. After the middle of 2000 the GJS values remain low. Hence,
comparing tick frequency behavior before and after mid 2000,
revealed that tick frequencies in the foreign exchange market after mid
2000 became increasingly similar to those in the first half of that year.
The similarity of the recent market appears to be the similarity in the
behavior of market participants all over the world as a result of the
development of information and communication technology.

\begin{table}[hbt]
\caption{Daily averages in April 2001, in millions of US dollars and
 percentages.}
\label{tab:share}
\begin{tabular}{lllll}
time zone & USD & EUR & JPY \\
\hline
Asian time zone & 373,179 (25.3\%) & 74,745 (12.2\%) & 160,384 (43.4\%) \\
\hline
European time zone & 806,997 (54.8\%) & 430,156 (70.3\%) & 137,731 (37.3\%) \\
\hline
American time zone & 292,563 (19.9\%) & 106,909 (17.5\%) & 71,448 (19.3\%) \\
\hline
Total & 1,472,739 (100\%) & 611,810 (100\%) & 369,563 (100\%) \\
\end{tabular}
\end{table}

According to the Triennial Central Bank Survey 2001~\cite{BIS:01}, 
the turnover by currency pairs in cross-border double-counting is
reported to be 30\% for USD/EUR, 20\% for USD/JPY, and 3\% for EUR/JPY. 
As shown in Table \ref{tab:share} by the share of currencies for various time zones, the USD and the EUR are the most actively traded 
in the European time zone and the JPY is actively traded
in both the Asian time zone and the European time zone. Table
\ref{tab:share} indicates that European market participants are active. 
Given that market participants in the European market primarily trade
with the EUR as an axis, it is predicted that the EUR/USD and the
EUR/JPY between Asia and Europe and Europe and America will behave
similarly in time, and that the dissimilarities among the EUR/USD,
USD/JPY, and EUR/JPY are likely to arise when the European market
opens. 

\begin{figure}[hbt]
\begin{center}
\includegraphics[scale=0.3]{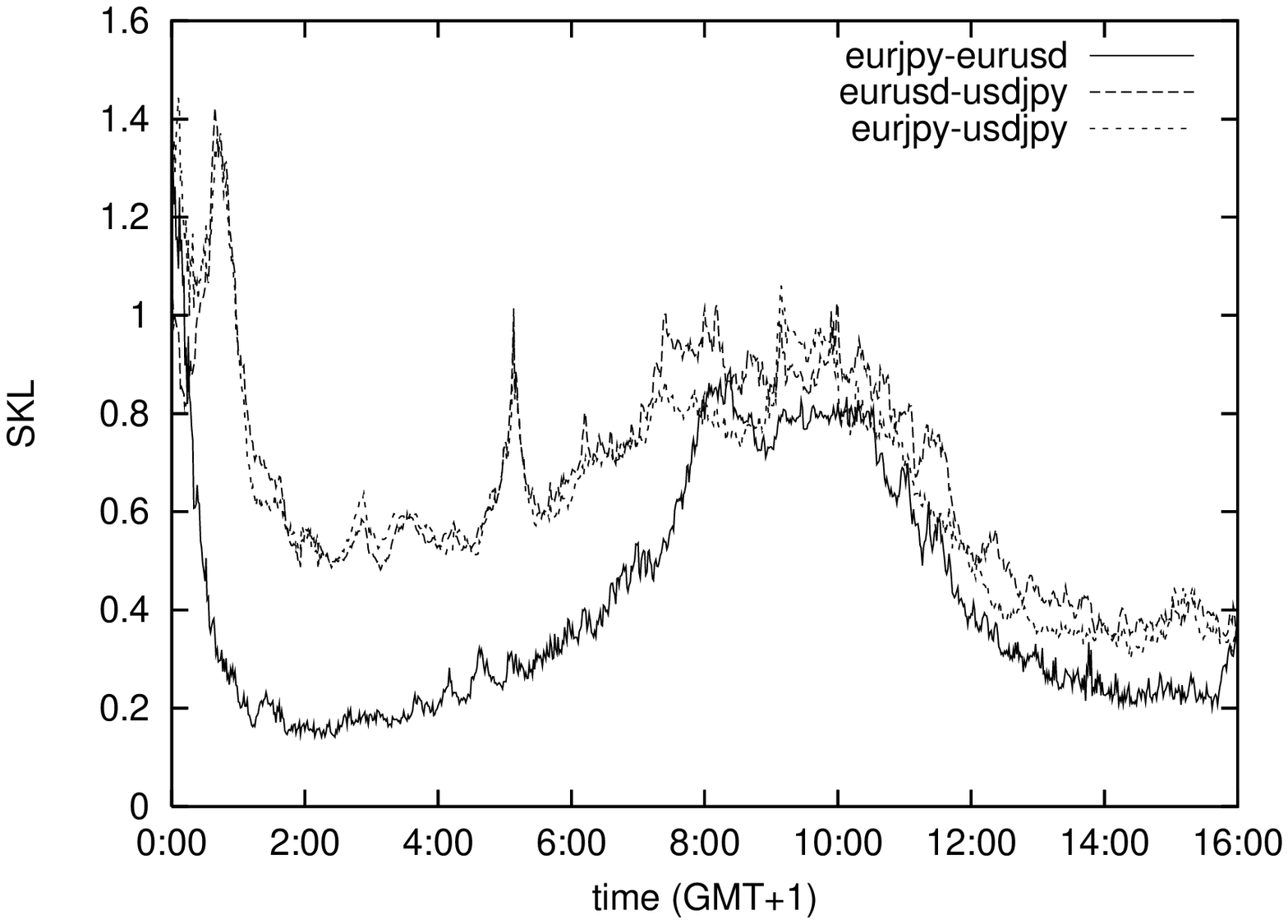}
\end{center}
\caption{The symmetric Kullback-Leibler divergence of spectrograms
 between the EUR/JPY and the EUR/USD (solid line), between the EUR/USD and the
 USD/JPY (long-dashed line), and between the EUR/JPY and the EUR/USD  
 (short-dashed line) on the 4th of September 2000.}
\label{fig:intraday}
\end{figure}

Both characteristics are found in the temporal development of the SKL,
as shown in Fig. \ref{fig:intraday}. The property whereby the spectral
distance is small at the time between the Asian time zone and the
European time zone, and between the European time zone and American
time zone, is also found in the GJS among 12 currency pairs
(Fig. \ref{fig:JS12}). It is thought that the market participants in each
time zone tend to exchange currencies with the market participants in
other time zones.

\section{Agent-based model}
\label{sec:agent-based-model}
Consider $N$ market participants who deal with $M$ currency pairs. The
$i$-th market participant perceives information $x_i(t)$ from the
environment. Based on this information the participant determines
his/her own investment attitude. Let $y_{ij}(t)$ denote the investment
attitude of the $i$-th market participant for the $j$-th currency
pair. The market participants are able to select his/her action from
among three investment attitudes (buying, selling, and waiting).

According to the Virginia Satir seven step model from perception to
action, which is one of psychological models of human~\cite{Weinberg:86},
an agent-based model of the foreign exchange market is considered. In the
Virginia Satir model, the process from perception to action is separated
into seven main steps: perception of information, interpretation, 
feeling, feeling about the feeling, defense, rule for acting, and action.

For simplicity, the $i$-th market participant perceives information
$x_i(t)$, which is evaluated as a scalar value. This information builds
momentum for the market participant to decides his/her investment
attitude. The market participant interprets the information and
determines his/her attitude on the basis of the interpretation. Since
the possibility of interpretation is very high and is dependent on time
and market participants, the uncertainty for the $i$-th agent to
interpret the information $x_i(t)$ a time $t$ is uniquely modeled by a
random variable $\xi_i(t)$. It is assumed that the result of the 
interpretation $x_i(t)+\xi_i(t)$ drives feeling, determining his/her
investment attitude. Furthermore, the feeling about the feeling examines the
validity of the feeling and drives his/her actions. In order to model
the feeling about the feeling, a multiplicative factor
$a_{ij}(t)$, which represents the feeling about the feeling of the $i$-th
market participant for the $j$-th currency pair, is introduced. If
$a_{ij}(t)$ is positive, then the feeling about the feeling supports the
feeling. If $a_{ij}(t)$ is negative, then the feeling about the feeling
refutes the feeling. The absolute value of $a_{ij}(t)$ represents the
intensity of the feeling about the feeling. Since the determination
depends on both the feeling and the feeling about the feeling, the
investment attitude is assumed to be determined from the value
$a_{ij}(t)(x_i(t)+\xi_i(t))$. If it is large, then the market participant
tends to make a buy decision. Contrarily, if it is small then the market
participant tends to make a sell decision. For simplicity, it is assumed
that a trading volume can be ignored.

The action is determined on the basis of the feeling about the
feeling of the market participant. Since the decision and action have strong nonlinearity, the action is determined with Granovetter type threshold
dynamics~\cite{Granovetter:78}. In order to separate three actions, at least two
thresholds are needed. Defining the threshold for the $i$-th market
participant to decide to buy the $j$-th currency pair 
as $\theta_{ij}^B(t)$ and to sell it as $\theta_{ij}^S(t)$
($\theta_{ij}^B(t) > \theta_{ij}^S(t)$), three investment attitudes 
(buying: 1, selling: -1, and waiting: 0) are determined by
\begin{equation}
y_{ij}(t)=
\left\{
\begin{array}{ll}
1 & (a_{ij}(t)(x_i(t)+\xi_i(t)) \geq \theta_{ij}^B(t)) \\
0 & (\theta_{ij}^B(t) < a_{ij}(t)(x_i(t) + \xi_i(t)) < \theta_{ij}^S(t)) \\
-1 & (a_{ij}(t)(x_i(t)+\xi_i(t)) \leq \theta_{ij}^S(t))
\end{array}
\right.
\end{equation}

Furthermore, it is assumed that the information is described as the
endogenous factor, the moving average of the log return over $T_{ij}(t)$, plus the
exogenous factor, $s_i(t)$:
\begin{equation}
x_i(t) = \sum_{k=1}^{M}C_{ik}(|\theta_{ik}^S(t)|,|\theta_{ik}^B(t)|)\frac{1}{T_{ik}(t)}\sum_{\tau=1}^{T_{ij}(t)}R_j(t-\tau\Delta t) + s_i(t), 
\label{eq:information}
\end{equation}
where $C_{ij}(|\theta_{ij}^S(t)|,|\theta_{ij}^B(t)|)$ represent the focal
points of the $i$-th market participant for the $j$-th currency pair. It
seems reasonable to assume that $C_{ij}(x,y)$ is a monotonically decreasing
function of $x$ and $y$.

The excess demand for the $j$-th currency pair, $N^{-1}\sum_{i=1}^N
 y_{ij}(t)$, drives the market price of the $j$-th currency
pair~\cite{McCauley:04}. To guarantee positive market prices, 
the following log return is chosen:
\begin{equation}
R_j(t) = \log S_j(t+\Delta t) - \log S_j(t),
\end{equation}
and define the log returns as the excess demand,
\begin{equation}
R_j(t) = \gamma N^{-1}\sum_{i=1}^N y_{ij}(t),
\label{eq:returns}
\end{equation}
where $\gamma$ is a positive constant to show the response of the
return to the excess demand. Furthermore, the tick frequency for the
$j$-th currency pair is defined as
\begin{equation}
A_j(t) = \frac{1}{\Delta t}\sum_{i=1}^N |y_{ij}(t)|.
\label{eq:tick-frequency}
\end{equation}

\begin{figure}[h]
\centering
\includegraphics[scale=0.5]{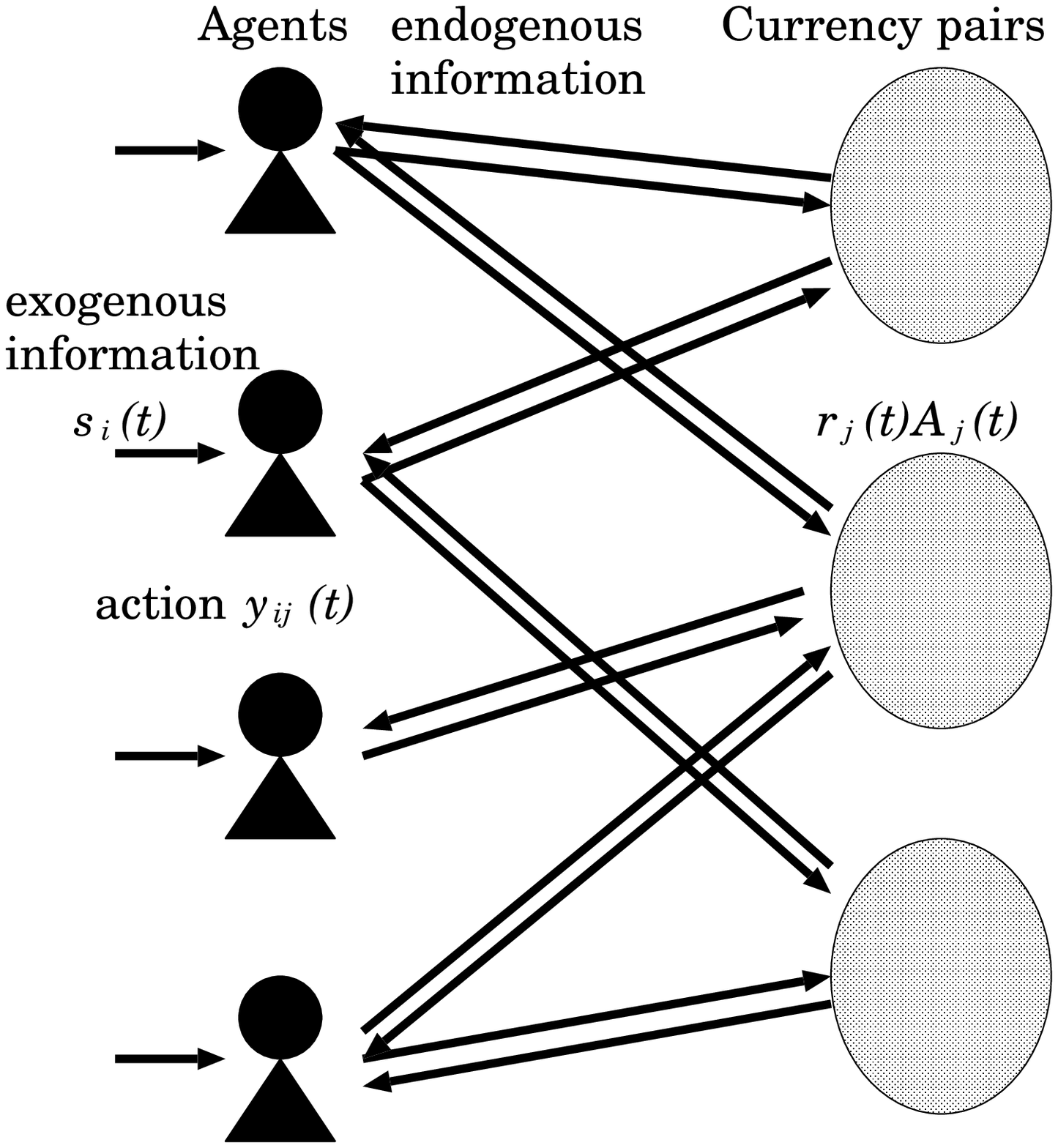}
\caption{A conceptual illustration of the agent model. $N$ market
 participants (agents) attend the market in order to exchange $M$
 currency pairs.}
\end{figure}

\section{Discussion}
\label{sec:discussion}
Since the market participants have a limitation due to finite time
and capacity to perceive the information and determine the investment
attitude, it is reasonable to assume that $T_{ij}(t)$ is finite and constant,
i.e., $T_{ij}(t)=T_{ij}$. Furthermore, for simplicity, it is assumed that
$\xi_i(t)$ is sampled from the independent Gaussian distribution,
\begin{equation}
p_{\xi}(\xi_i) = \frac{1}{\sqrt{2\pi}\sigma}\exp\Bigl(-\frac{\xi_i^2}{2\sigma^2}\Bigr),
\end{equation}
where $\sigma (>0)$ represents the standard deviation of the uncertainty of
the interpretation. Then, $y_{ij}(t)$ are random variables that take 1, 0,
and -1 with the probabilities $Q_{ij}(y|x_i(t))$: 
\begin{eqnarray}
Q_{ij}(1|x_i(t)) &=&
 \frac{1}{2}\mbox{erfc}\Bigl(\frac{\theta_{ij}^S(t)/a_{ij}(t)-x_i(t)}
 {\sqrt{2}\sigma}\Bigr), 
\label{eq:Pp}
\\
Q_{ij}(-1|x_i(t)) &=& 
 \frac{1}{2}\mbox{erfc}\Bigl(\frac{\theta_{ij}^B(t)/a_{ij}(t)+x_i(t)}
 {\sqrt{2}\sigma}\Bigr),
\label{eq:Pm}
 \\ 
Q_{ij}(0|x_i(t)) &=& 1 - Q_{ij}(1|x_i(t)) - Q_{ij}(-1|x_i(t)),
\label{eq:Pz}
\end{eqnarray}
where $x_i(t) = \sum_{k=1}^{M}
C_{ik}(|\theta_{ik}^S|,|\theta_{ik}^B|)\frac{1}{T_{ik}}\sum_{\tau=1}^{T_{ik}}
R_k(t-\tau\Delta t)+s_i(t)$. 
$\mbox{erfc}(x)$ is the complementary error function defined as
\begin{equation}
\mbox{erfc}(x) = \frac{2}{\sqrt{\pi}}\int_{x}^{\infty}e^{-u^2}du.
\end{equation}
From Eqs. (\ref{eq:Pp}), (\ref{eq:Pm}), and (\ref{eq:Pz}), we obtain
\begin{eqnarray}
\langle y_{ij}(t) \rangle &=& Q_{ij}(1|x_i(t)) - Q_{ij}(-1|x_i(t)),
\label{eq:en-y}
\\
\langle |y_{ij}(t)| \rangle &=& Q_{ij}(1|x_i(t)) + Q_{ij}(-1|x_i(t)).
\label{eq:en-ay}
\end{eqnarray}
From Eqs. (\ref{eq:returns}) and (\ref{eq:tick-frequency}), the
ensemble averages of $R_j(t)$ and $A_j(t)$ are approximated by 
\begin{eqnarray}
\langle R_j(t) \rangle &=& \gamma N^{-1} \sum_{i=1}^N \langle y_{ij}(t)
 \rangle, 
\label{eq:en-returns-1}
\\
\langle A_j(t) \rangle &=& \frac{1}{\Delta t} \sum_{i=1}^N \langle
|y_{ij}(t)| \rangle.
\label{eq:en-tick-frequency-1}
\end{eqnarray}
Therefore, the substitution of Eqs. (\ref{eq:en-y}) and (\ref{eq:en-ay}) into
Eqs. (\ref{eq:en-returns-1}) and (\ref{eq:en-tick-frequency-1}) yields
\begin{eqnarray}
\langle R_j (t) \rangle &=&
\gamma N^{-1}
\sum_{i=1}^{N}f\Bigl(x_i(t);\theta_{ij}^S(t)/a_{ij}(t),\theta_{ij}^B(t)/a_{ij}(t)\Bigr), 
\label{eq:en-returns}
\\
\langle A_j (t) \rangle &=&
\frac{1}{\Delta t}
\sum_{i=1}^{N}g\Bigl(x_i(t);\theta_{ij}^S(t)/a_{ij}(t),\theta_{ij}^B(t)/a_{ij}(t)\Bigr),
\label{eq:en-tick-frequency}
\end{eqnarray}
where
\begin{eqnarray}
f(x;a,b) &=& \frac{1}{2}\mbox{erfc}\Bigl(\frac{a-x}{\sqrt{2}\sigma}\Bigr)
 - \frac{1}{2}\mbox{erfc}\Bigl(\frac{b+x}{\sqrt{2}\sigma}\Bigr), \\
g(x;a,b) &=& \frac{1}{2}\mbox{erfc}\Bigl(\frac{a-x}{\sqrt{2}\sigma}\Bigr)
 + \frac{1}{2}\mbox{erfc}\Bigl(\frac{b+x}{\sqrt{2}\sigma}\Bigr).
\end{eqnarray}

Here, the agent variation is assumed to be constant with time during
the observation period. Then, $\theta_{ij}^{S/B}(t)$ and $a_{ij}(t)$ are
slowly varying functions of $t$ and the assumption that they are
constant is reasonable, so that $\theta_{ij}^{S/B}(t) = \theta_{ij}^{S/B}$
and $a_{ij}(t) = a_{ij}$. Furthermore, if $R_j(t)=\langle R_j(t) \rangle
+ \eta_j(t)$ and $A_j(t) = \langle A_j(t) \rangle + \epsilon_j(t)$,
where $\eta_j(t)$ and $\epsilon_j(t)$ are identically and independently
distributed noises then
\begin{eqnarray}
R_j(t) &=& \gamma N^{-1}
 \sum_{i=1}^{N}f\Bigl(x_i(t);\theta_{ij}^S/a_{ij},\theta_{ij}^B/a_{ij}\Bigr)+\eta_j(t), 
\label{eq:return1}
\\ 
A_j(t) &=& \frac{1}{\Delta t}
 \sum_{i=1}^{N}g\Bigl(x_i(t);\theta_{ij}^S/a_{ij},\theta_{ij}^B/a_{ij}\Bigr) +
 \epsilon_j(t).
\label{eq:tick-frequency1}
\end{eqnarray}
The power spectral density for $A_j(t)$ is defined as
\begin{equation}
P_j(f;\{\theta_{ij}^{S}/a_{ij}\},\{\theta_{ij}^{B}/a_{ij}\}) =
 \Bigl|\sum_{t=-\infty}^{\infty} A_j(t) e^{-2\pi \mbox{i} t f}\Bigr|^2 
\quad (-\frac{1}{2}<f<\frac{1}{2}).
\label{eq:power-spectral-of-A}
\end{equation}

Furthermore, the Taylor expansion of Eq. (\ref{eq:tick-frequency1}) is
written as
\begin{equation}
A_j(t) = \frac{1}{\Delta t}\sum_{i=1}^{N}\Bigl\{g\bigl(0;\theta_{ij}^S/a_{ij},
\theta_{ij}^B/a_{ij}\bigr) + g'\bigl(0;\theta_{ij}^S/a_{ij},
\theta_{ij}^B/a_{ij}\bigr)x_i(t) +O(x_i(t)^3) \Bigr\}+ \epsilon_j, 
\end{equation}
where $g'(0;a,b)=\sqrt{(2/\pi)}(1/\sigma)\{\exp[-a^2/2\sigma^2]-\exp[-b^2/2\sigma^2]\}$. If the information that the $i$-th agent perceives is weak, so that,
$|x_i(t)|$ is small, then terms of order higher than 
the first order can be neglected, and we obtain
\begin{equation}
A_j(t) = \frac{1}{\Delta t}\sum_{i=1}^{N}\Bigl\{g\bigl(0;\theta_{ij}^S/a_{ij},
\theta_{ij}^B/a_{ij}\bigr)+ g'\bigl(0;\theta_{ij}^S/a_{ij},
\theta_{ij}^B/a_{ij}\bigr)x_i(t)\Bigr\} + \epsilon_j.
\end{equation}
Then, Eq. (\ref{eq:power-spectral-of-A}) is given by
\begin{equation}
P_j(f;\{\theta_{ij}^{S}/a_{ij}\},\{\theta_{ij}^{B}/a_{ij}\})=
\Bigl|\tilde{g}_j + \sum_{i=1}^{N}\hat{g}_{ij}\hat{x}_i(n)+\hat{\epsilon}_j(f) 
\Bigr|^2,
\end{equation}
where $\tilde{g}_j = (1/\Delta t)
\sum_{i=1}^{N}g(0;\theta_{ij}^S/a_{ij},\theta_{ij}^B/a_{ij})\sum_{t=-\infty}^{\infty}\exp[-2\pi\mbox{i}tf]$,
$\hat{g}_{ij}=(1/\Delta t)g'(0;\theta_{ij}^S/a_{ij}, \theta_{ij}^B/a_{ij})$,
$\hat{x}_i(n)=\sum_{t=-\infty}^{\infty} x_i(t) \exp[-2\pi \mbox{i} t f]$, 
and $\hat{\epsilon}_j(f)=\sum_{t=-\infty}^{\infty} \epsilon_j(t)\exp[-2\pi
\mbox{i} t f]$, and the power normalized spectral density is
calculated as 
\begin{equation}
p_j(f;\{\theta_{ij}^{S}/a_{ij}\},\{\theta_{ij}^{B}/a_{ij}\}) =
\left\{
\begin{array}{ll}
 \frac{P_j(f;\{\theta_{ij}^{S}/a_{ij}\},\{\theta_{ij}^{B}/a_{ij}\})}{\int_{-1/2}^{1/2}P_j(f;\{\theta_{ij}^{S}/a_{ij}\},\{\theta_{ij}^{B}/a_{ij}\})df}
  & (f \neq 0) \\
0 & (f = 0)
\end{array}
\right..
\end{equation}

Since $p_j(f;\{\theta_{ij}^{S}/a_{ij}\},\{\theta_{ij}^{B}/a_{ij}\})$ 
are functions in terms of $\theta_{ij}^S/a_{ij}$, 
$\theta_{ij}^B/a_{ij}$ for any $i$, the spectral distance
of the tick frequency between the $l$-th currency pair and the $m$-th
currency pair,
\begin{eqnarray}
\nonumber
K_{lm}(\{\theta_{il}^{S}/a_{il}\},\{\theta_{il}^{B}/a_{il}\},\{\theta_{im}^{S}/a_{im}\},\{\theta_{im}^{B}/a_{im}\}) = \\
 \int_{-1/2}^{1/2}p_l(f;\{\theta_{il}^{S}/a_{il}\},\{\theta_{il}^{B}/a_{il}\})\log\frac{p_l(f;\{\theta_{il}^{S}/a_{il}\},\{\theta_{il}^{B}/a_{il}\})}{p_m(f;\{\theta_{im}^{S}/a_{im}\},\{\theta_{im}^{B}/a_{im}\})}df, 
\label{eq:KL}
\end{eqnarray} 
are functions of $\theta_{ij}^S/a_{ij}$, $\theta_{ij}^B/a_{ij}$ for 
any $i$ and $j=l,m$. If $\theta_{il}^S/a_{il} = \theta_{im}^S/a_{im}$, 
$\theta_{il}^B/a_{il} = \theta_{im}^B/a_{im}$ for any $i$, then
$\hat{g}_{il} = \hat{g}_{im}$ for any $i$ and $K_{lm}$ is equal to zero. 
Hence, the differences between $\theta_{il}^S/a_{il}$ and
$\theta_{im}^S/a_{im}$, and between $\theta_{il}^B/a_{il}$ and
$\theta_{im}^B/a_{im}$, reflect $K_{lm}$. Since $\theta_{ij}^S$ and
$\theta_{ij}^B$ represent the $i$-th agent's decision and perception,
respectively, for the $j$-th currency pairs, $a_{ij}$ represents the
feeling about the feeling of the $i$-th agent for the $j$-th currency
pair and $K_{lm}$ is associated with the behavioral similarity of the
market participants who exchange the $l$th-currency pair and the
$m$th-currency pair. Namely, the similarity of the tick frequency
between the $l$th-currency pair and the $m$th-currency pair is
equivalent to the similarity of the behavior (perception and decision)
for which the market participants exchange the $l$th-currency pair and
the $m$th-currency pair. Therefore, these quantities can characterize
the behavioral structure of the participants in the market.

Moreover, the SKL
\begin{equation}
J_{lm} = \frac{1}{2}(K_{lm} +K_{ml})
\label{eq:SKL}
\end{equation}
can be described using the normalized autocorrelation
functions and the cepstral coefficients~\cite{Veldhuis:03}. Let $r_l(t)$
denote the normalized autocorrelation function and $c_l(t)$ the cepstral
coefficients:
\begin{eqnarray}
p_l(f) &=& \sum_{t=-\infty}^{\infty}r_l(t)e^{-2\pi \mbox{i}t f}, 
\label{eq:autocorrelation}
\\
\log p_l(f) &=& \sum_{t=-\infty}^{\infty}c_l(t) e^{-2\pi
 \mbox{i}t f}.
\label{eq:cepstrum}
\end{eqnarray}
By substituting Eqs. (\ref{eq:autocorrelation}) and (\ref{eq:cepstrum})
into Eq. (\ref{eq:KL}) and using (\ref{eq:SKL}), we obtain
\begin{equation}
J_{lm} = \sum_{t=-\infty}^{\infty}(r_l(t)-r_m(t))(c_l(t)-c_m(t)).
\end{equation}
$J_{lm}$ is also associated with the similarity between the normalized
autocorrelation functions and between the cepstral coefficients. 

If the spectral distance is measured by the JS
\begin{equation}
JS_{lm}^\pi = H(\pi_1p_i+\pi_2p_j)-\pi_1H(p_i)-\pi_2H(p_j),
\end{equation}
where $\pi_1,\pi_2>0$ and $\pi_1+\pi_2=1$, are the weights of the
two power normalized spectra, and $H$ is the Shannon entropy function,
$H(p)=-\int_{-1/2}^{1/2} p(f) \log p(f) df$,
then
\begin{equation}
JS_{lm}^\pi = \sum_{t=-\infty}^{\infty}\Bigl(\pi_i r_l(t)c_l(t) + 
\pi_j r_m(t)c_m(t) - \tilde{r}_{lm}(t)\tilde{c}_{lm}(t) \Bigr),
\end{equation}
where $\tilde{r}_{lm}(t)$ and $\tilde{c}_{lm}(t)$ are an 
average normalized autocorrelation coefficient and averaged cepstral
coefficient, respectively, defined as
\begin{eqnarray}
\tilde{r}_{lm}(t) &=&  \pi_i r_i(t)+\pi_j r_j(t), \\
\tilde{c}_{lm}(t) &=&
 \frac{1}{2\pi}\int_{-1/2}^{1/2}\log\bigl(\pi_ip_i(f)+\pi_jp_j(f)\bigr)e^{\mbox{i}tf}df. 
\end{eqnarray}
The JS allows us to measure the spectral distance with {\it a priori}
probability for $p_i(f)$. Furthermore, the GJS are expressed as 
\begin{equation}
JS^\pi(p_1,p_2,\ldots,p_M) = \sum_{t=-\infty}^{\infty}\Bigl(
\sum_{i=1}^M\pi_i r_i(t)c_i(t) - \tilde{r}(t)\tilde{c}(t)\Bigr),
\end{equation}
by using an averaged normalized autocorrelation coefficient and average
cepstral coefficient defined as
\begin{eqnarray}
\tilde{r}(t) &=&  \sum_{i=1}^M \pi_i r_i(t), \\
\tilde{c}(t) &=&
 \frac{1}{2\pi}\int_{-1/2}^{1/2}\log\Bigl(\sum_{i=1}^M\pi_ip_i(f)\Bigr)e^{\mbox{i}tf}df. 
\end{eqnarray}
The SKL, the JS, and the GJS as a measure to quantify the spectral
distance are associated with autocorrelation coefficients and cepstral
coefficients. 

As shown in Section \ref{sec:spectral-analysis}, the SKL and the JS
among the currency pairs vary temporally and similar pairs and
dissimilar pairs exist. These temporal structure variations of the
spectral distance appears to capture and characterize the behavioral
parameters of the market participants. The perception and decision of
the market participants who deal with these currency pairs vary
temporally, but several types of similar and dissimilar patterns of
perception and decision exist. 

As shown in Figs. \ref{fig:JS12} and \ref{fig:intraday} the fact that
the GJS/SKL is lower in the time at inter-area than in the time at
intra-area implies that agents parameters become more similar in the
time inter-area than in the time at intra-area. The reason is thought to
be because market participants are able to exchange a wide variety of
currencies in the time at intra-area and attempt cross
arbitration. Since several market participants have a tendency to trade
with the same awareness and strategies the microscopic parameters of the
them may be in the narrow range.
 
Thus, it is possible to compare the market participants' parameters for
perception and decision by analyzing the tick frequency.

\section{Conclusions}
\label{sec:conclusions}
The tick frequency of the foreign exchange market for 12
currency pairs (EUR/CHF, EUR/GBP, EUR/JPY, EUR/NOK, EUR/SEK, EUR/USD,
NZD/USD, USD/CAD, USD/CHF, USD/JPY, USD/NOK, and USD/SEK) was
investigated. By utilizing the spectral distance based on the
Kullback-Leibler divergence between two normalized spectrograms and the
Jensen-Shannon divergence among them, the behavioral structure of market
participants in the foreign exchange market was found to vary
dynamically, and the present markets were found to be more similar than
past markets. In order to understand the meaning of the similarity
between two currency pairs, the agent-based model in which $N$ market
participants exchange $M$ currency pairs was considered. The spectral
distance for the tick frequency was concluded to characterize the
similarity of the market participants' behavioral parameters.

Analyzing the tick frequency, as well as the prices or rates, will provide 
deep insights into the behaviors of market participants in financial markets 
from the perspectives of both finance and physics.

\section*{Acknowledgement}
This study was supported by a Grant-in-Aid for Scientific Research (\# 17760067) from the Ministry of Education, Culture, Sports, Science and Technology.

\end{document}